\documentstyle[preprint,eqsecnum,aps,epsf,floats,tighten]{revtex} 

\begin{document}

\preprint{\tighten \vbox{\hbox{DO-TH 97/27} 
		\hbox{hep-ph/9801230} \hbox{} }}

\title{Analysis of hadronic invariant mass spectrum in inclusive 
charmless semileptonic $B$ decays}

\author{Changhao Jin}

\address{Institut f\"ur Physik, Universit\"at Dortmund\\
D--44221 Dortmund, Germany}

\maketitle

{\tighten
\begin{abstract}%
We make an analysis of the hadronic invariant mass spectrum in inclusive
charmless semileptonic $B$ meson decays in a QCD-based approach.
The decay width is studied as a function of the invariant mass cut.
We examine their sensitivities to the parameters of the theory. 
The theoretical uncertainties in the determination of $|V_{ub}|$ from
the hadronic invariant mass spectrum are investigated.  A strategy
for improving the theoretical accuracy in the value of $|V_{ub}|$
is described. 
 
\end{abstract}
}

\newpage

\section{Introduction}
The Cabibbo-Kobayashi-Maskawa (CKM) matrix element $V_{ub}$ is a fundemental  
parameter in the standard model of particle physics, 
which is not yet known with sufficient precision. A precise determination 
of it is very important for testing the standard model and for studying the 
origin of CP violation and quark masses. 

Until recently, the sole evidence for 
charmless semileptonic $B$ meson decays and for a non-zero value of $V_{ub}$,
which occurs as a multiplicative factor in the decay amplitude, 
was reported from the inclusive charged
lepton (electron or muon) energy spectra \cite{cleo1,argus}.
Since the rate of the charmless semileptonic $B$ decay is very small, the
main experimental challenge in these analyses is the suppression of 
backgrounds.  The charged lepton from the inclusive charmless semileptonic  
decay $B\to X_u\ell\nu$ ($\ell=e$ or $\mu$) can be more energetic than that 
from the dominant semileptonic $b\to c$ decay; for a $B$ meson at rest, the 
spectrum in the endpoint region $(M/2)(1-M_D^2/M^2) < E_\ell\leq
(M/2)(1-M_\pi^2/M^2)$ ($M$ being the mass of the $B$ meson, $M_D$ the mass of 
the $D$ meson, and $M_\pi$ the mass of the pion) results only from the 
$b\to u$ decay. This simple kinematic cut has been used to obtain the first 
sign of semileptonic
$b\to u$ transitions\footnote{A new inclusive analysis was presented 
by ALEPH which is the first evidence for semileptonic $b\to u$ transitions
in $b$-hadrons produced at LEP\cite{aleph}.}
and determine $|V_{ub}|$ from the 
inclusive charged lepton energy spectra \cite{richman}. 
The theoretical description of the decays requires solving QCD in the
nonperturbative regime.
The accuracy of this inclusive determination of $|V_{ub}|$ is limited by model 
dependence, but
recent work with a QCD-based approach shows promise of improving the 
situation \cite{jp1}.     

Recently, the CLEO Collaboration has
studied the exclusive charmless semileptonic $B$ decays to $\pi\ell\nu$, 
$\rho\ell\nu$, and $\omega\ell\nu$ \cite{cleo2}. 
This is achieved through the reconstruction of semileptonic decays involving a
$b\to u$ transition. The final states are reconstructed using an observed
hadron, an $e$ or $\mu$, and a neutrino. 
Using the hermeticity of the CLEO detector,
they reconstruct the neutrino by inferring its four-momentum from 
the missing energy and missing momentum in 
each event.  With this neutrino reconstruction
technique, they have measured the branching fractions for
$B^0\to\pi^-\ell^+\nu$ and $B^0\to\rho^-\ell^+\nu$  
and have extracted $|V_{ub}|$ from this study.
Current determination of $|V_{ub}|$ by this exclusive method has large model 
dependence \cite{cleo2,gib}. 
One should notice, however, recent significant theoretical progress in 
describing exclusive semileptonic $b\to u$ decays \cite{rev}.   

Various other methods to determine $|V_{ub}|$ have been proposed.
These efforts are obviously necessary in order to find the most precise way
to extract this important parameter.  At least, different
methods and analyses including two existing types of measurements  
lend confidence and indicate limitations. 
Among them it has been proposed \cite{bar}
long time ago that the hadronic invariant mass spectrum in the inclusive 
semileptonic $B$ decays may be useful for the determination of $|V_{ub}|$.
For a $B$ meson at rest,
events with a kinematic cut on the hadronic invariant mass in the final
states, $M_X<M_D$, 
come only from the $b\to u$ transition (the full $b\to u$ kinematic range is
$M_\pi\leq M_X\leq M$) and are thus separated from those due to 
the $b\to c$ transition. Using this cut we have access to a 
much larger region of phase space than with the cut on $E_\ell$.
About $90\%$ of the $b\to u$ events survive this cut in contrast 
to the charged lepton energy cut, with which only about $10\%$ of the $b\to u$
events survive. Hence the theoretical calculation of the decay rate is 
expected to be more reliable with the hadronic invariant mass cut.
Recent experimental work on the neutrino reconstruction technique, which
has been used by CLEO in the exclusive charmless semileptonic $B$ decay
measurement \cite{cleo2},  
shows promise of measuring the hadronic invariant mass spectrum.
A number of recent papers \cite{sev} have renewed discussions on such
possibility of determining $|V_{ub}|$. On the other hand, studies of the 
hadronic invariant mass spectra will also provide a testing ground for our 
understanding of the decay dynamics.

In this paper we will study the hadronic 
invariant mass spectrum and the decay width with a hadronic invariant mass
cut in inclusive charmless semileptonic $B$ meson decays 
in a QCD-based approach \cite{jp,jin,jp1}. 
This approach describes inclusive
semileptonic $B$ meson decays by making use of the
light-cone expansion and the heavy quark effective theory (HQET) and has been 
recently extended to inclusive semileptonic decays of an unpolarized 
$b$-flavored hadron \cite{jin1} . 
The light-cone dominance attributes
the nonperturbative QCD effects on the underlying weak decays to 
a single distribution function.
The sum rules derived from operator product expansion and heavy quark effective
theory constrain the distribution function considerably.
The interplay between nonperturbative and perturbative QCD effects has been
accounted for and a coherent treatment of both
has been formulated. This approach
has been tested on inclusive semileptonic decays of $B$ mesons.
The resulting lepton energy spectrum is in good agreement with the experimental
data \cite{jp1}.
The calculation of the semileptonic decay width of the $B$ meson has 
shown \cite{jin}
that the kinematically enhanced nonperturbative QCD contributions play 
numerically an important role and $|V_{cb}|$ has then been determined,
which is consistent with the exclusive measurements \cite{drell}. 
It has been shown \cite{jp1} 
that an improved precision in $|V_{ub}|$ can be gained from the 
lepton energy spectra with a controlled theoretical uncertainty. 
In this work we turn our attention to investigate the 
theoretical accuracy of the alternative determination of $|V_{ub}|$,
i.e., from the hadronic invariant mass spectrum.

Section II presents the theoretical methods to calculate the charmless
hadronic invariant mass spectrum. We include both the bound-state
effects and the QCD radiative corrections. In Sec.~III we
examine the sensitivities of the spectrum to the parameters of the theory.
The effect of the QCD radiative corrections is also discussed.
The theoretical uncertainties in the calculation of the semileptonic 
decay width to a given maximum hadronic invariant mass for $B\to X_u\ell
\nu$ are studied in Sec.~IV.
We estimate the theoretical errors in the determination of $|V_{ub}|$ from
the hadronic invariant mass spectrum. Finally, concluding remarks and
a discussion are given in Sec.~V.

\section{Theoretical Method}

The approach has been described in [11-13,5], which we refer
to for a more complete exposition of the theory.
We begin by briefly reviewing the methods relevant to this analysis. 
Because of the light-cone dominance,
the nonperturbative QCD effects on the inclusive semileptonic $B$ meson decays
can be disentangled and encoded in a distribution function:
\begin{equation}
f(\xi)=\frac{1}{4\pi M^2}\int d(y\cdot P)e^{i\xi y\cdot P}
\langle B\left| \bar b(0)/ \mkern -12mu P(1-\gamma_5)b(y)
\right|B\rangle \
|_{y^2=0} ,
\label{eq:dlight}
\end{equation}
where $P$ denotes the four-momentum of the $B$ meson.
Several important properties of the distribution function were derived from
QCD. Due to current conservation, it is exactly normalized to unity 
with a support $0\leq\xi\leq 1$:
\begin{equation}
\int^1_0 d\xi f(\xi) =1.
\end{equation}
It obeys positivity.  It contains the free quark decay as a
limiting case with $f(\xi)=\delta(\xi-m_b/M)$, where $m_b$ labels the $b$
quark mass. In addition, two sum rules for the distribution function
can be obtained by using the operator product expansion and the HQET 
method \cite{hqe}.
These two sum rules determine the mean value $\mu$ and the variance
$\sigma^2$ of the distribution function,
which characterize the position of the maximum and the width of 
it, respectively:
\begin{equation}
\mu \equiv \int_0^1 d\xi\xi f(\xi) = \frac{m_b}{M} (1+E_b)\, ,
\label{eq:sum1}
\end{equation}
\begin{equation}
\sigma^2 \equiv \int_0^1 d\xi(\xi-\mu)^2 f(\xi) = 
\frac{m_b^2}{M^2} \left( \frac{2K_b}{3} - E_b^2 \right)\, ,
\label{eq:sum2}
\end{equation}
where 
\begin{equation}
G_b= -\frac{1}{2M} \left\langle B\left |\bar{h}_v
\frac{g_sG_{\alpha\beta}
\sigma^{\alpha\beta}}{4m_b^2} h_v\right |B\right\rangle\,,
\end{equation}
\begin{equation}
K_b= -\frac{1}{2M} \left\langle B\left |\bar{h}_v\,
  \frac{(iD)^2}{2m_b^2}\, h_v\right |B \right\rangle \, ,
\end{equation}    
with $E_b=G_b+K_b$. The first matrix element $G_b$
measures the chromomagnetic energy due to the spin coupling between 
the $b$ quark and the light constituents in the $B$ meson and is determined 
by the mass splitting between $B^*$ and $B$ mesons.
For the observed difference $M_{B^*}-M_B = 0.046$ GeV \cite{PDG}, one gets 
\begin{equation}
m_bG_b = -\frac{3}{4}(M_{B^*}-M_B)= -0.034\,\, {\rm GeV}\, .
\end{equation} 
The second matrix element $K_b$ 
measures the kinetic energy of the $b$ quark in the B meson.
A calculation using QCD sum rules yields \cite{ball}
\begin{equation}
2m_b^2K_b=0.5\pm 0.2\,\, {\rm GeV}^2\, .
\end{equation}
Taking $m_b=4.9\pm 0.2$ GeV, the mean value and the variance of the 
distribution function are estimated to be
\begin{equation}
 \mu = 0.93 \pm 0.04 \, ,
\label{eq:constrain1}
\end{equation}
\begin{equation}
\sigma^2 = 0.006 \pm 0.002\, .
\label{eq:constrain2}
\end{equation}
A non-zero value for $\sigma^2$ indicates the deviation from
$f(\xi)=\delta(\xi-m_b/M)$, i.e., the free quark decay, and is a measure
of the amount of the nonperturbative QCD contribution.

To obtain the hadronic invariant mass distribution $d\Gamma/dM_X$, it is
technically convenient to start from the double differential decay rate
$d^2\Gamma/dq^2dq^0$, since the momentum transfer to the lepton pair,
$q$, is a quantity which has no distinction between the hadron level and 
the quark level. 
Following \cite{jp1}, one can account for the nonperturbative and perturbative 
QCD effects on $d^2\Gamma/dq^2dq^0$ in a coherent way through the convolution 
of the quark-level decay rate and the distribution function.  
This double differential decay rate
is then straightforwardly converted to $d\Gamma/dM_X$. In the $B$ meson rest
frame, the hadronic invariant mass spectrum in the inclusive
charmless semileptonic $B$ decay is finally given by  
\begin{equation}
\frac{d\Gamma}{dM_X}=\frac{M_X}{M}\int_0^{(M-M_X)^2} dq^2
\int^1_{\frac{q^0+|{\bf q}|}{M}} d\xi\, f(\xi)\Bigg [\frac{d^2\Gamma_b}
{dq^2dq^0}\Bigg ]_{p_b=\xi P} ,
\label{eq:spec}
\end{equation}
where $p_b$ denotes the four-momentum of the $b$ quark.
The detailed expressions of the quark-level double differential decay 
rate $d^2\Gamma_b/dq^2dq^0$ including the
perturbative QCD corrections can be found in \cite{greub}. 
The masses of the leptons and the $u$ quark have been neglected.
To illustrate the effect of the radiative QCD corrections, it is instructive
to retain only the terms which do not arise from such gluon radiation.
Subtracting the radiative QCD corrections, 
Eq.~(\ref{eq:spec}) then reduces to 
\begin{equation}
\frac{d\Gamma}{dM_X}=\frac{G_F^2M_X|V_{ub}|^2}{24\pi^3}
\int_0^{(M-M_X)^2} dq^2\, |{\bf q}|\Bigg [f(\xi_+)\Bigg (\xi_+^2+
\frac{2q^2}{M^2}\Bigg )+(\xi_+\to\xi_-)\Bigg ] ,
\label{eq:spec1}
\end{equation}            
where 
\begin{equation}
\xi_\pm=\frac{q^0\pm|{\bf q}|}{M}.
\label{eq:pm}
\end{equation}

By integrating Eq.~(\ref{eq:spec}) over the hadronic invariant mass $M_X$, 
we obtain the decay width for the $B\to X_u\ell\nu$ decay to a given maximum 
hadronic invariant mass $M_X^{\rm cut}$ defined by
\begin{equation}
\Gamma(M_X^{\rm cut})=\int_0^{M_X^{\rm cut}} dM_X \frac{d\Gamma}{dM_X}.
\label{eq:partial}
\end{equation}

The known properties of the distribution function stated above improve
the theoretical accuracy remarkably. However, since the distribution function
has not yet been completely determined in QCD, for pratical calculations
we shall adopt the following 
parametrization \cite{jin} of the distribution function which incorporates
all known properties:
\begin{equation}
f(\xi)=N\frac{\xi (1-\xi)^\alpha}{[(\xi-a)^2+b^2]^\beta}\theta(\xi)
\theta(1-\xi) \ ,
\label{eq:ansatz}
\end{equation}
where $a, b, \alpha$, and $\beta$ are four parameters, which are constrained
by the sum rules (\ref{eq:sum1}) and (\ref{eq:sum2}), and $N$ is the 
normalization constant. 
For $\alpha=\beta=1$, Eq.(\ref{eq:ansatz}) becomes the ansatz of \cite{jp}.
In the following, $\alpha=\beta=1$ is preset unless explicitly 
stated\footnote{In general, the parameters 
$\alpha$ and $\beta$ need not be integer.}.
  
\section{Charmless Hadronic Invariant Mass Spectrum}

We proceed to examine the sensitivity of the charmless hadronic invariant 
mass spectrum to the parameters of the theory. 
We calculate the hadronic invariant mass spectrum in the inclusive charmless 
semileptonic $B$ meson decay for a $B$ meson at rest by use of 
Eqs.~(\ref{eq:spec}) and (\ref{eq:ansatz}). The input parameters are the strong
coupling constant $\alpha_s$ and the parameters which arise in the 
distribution function.

\begin{figure}[t]
\centerline{\epsfysize=9truecm \epsfbox{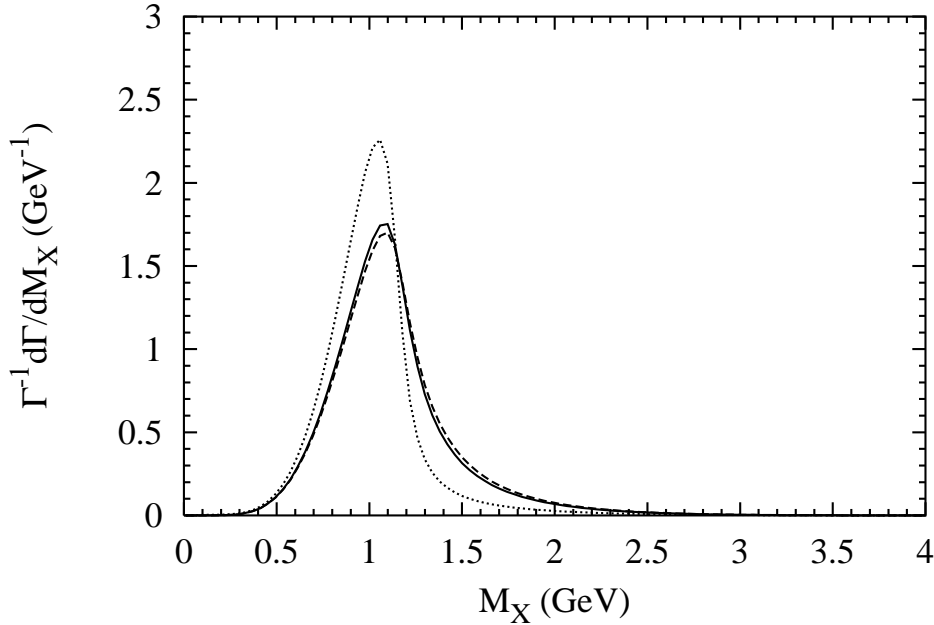}}
\tighten{
\caption[tau1]{The charmless hadronic invariant
mass spectrum with (solid curve: $\alpha_s=0.25$, 
dashed curve: $\alpha_s=0.30$) 
and without (dotted curve) radiative QCD corrections. 
The mean value and the variance of the distribution function are
kept fixed to be $\mu=0.93$ and $\sigma^2=0.006$.} 
\label{fig:mx_u1} }
\end{figure}

The effects of the radiative QCD corrections to the 
spectrum are demonstrated in Fig.~\ref{fig:mx_u1}. We calculate the spectra
using two different values of the strong coupling constant: $\alpha_s=0.25$
(solid curve) and $\alpha_s=0.30$ (dashed curve). The spectrum appears to be 
insensitive to the value of $\alpha_s$, varied within a reasonable range.  
For comparison, the spectrum without the radiative QCD
corrections (dotted curve) is also shown, 
calculated by use of Eqs.~(\ref{eq:spec1}) and
(\ref{eq:ansatz}). It is evident that the
charmless hadronic invariant mass spectrum receives large radiative QCD
corrections. This is easily understood because gluon bremsstrahlung strongly
affects the hadronic final states.

\begin{figure}[t]
\centerline{\epsfysize=9truecm \epsfbox{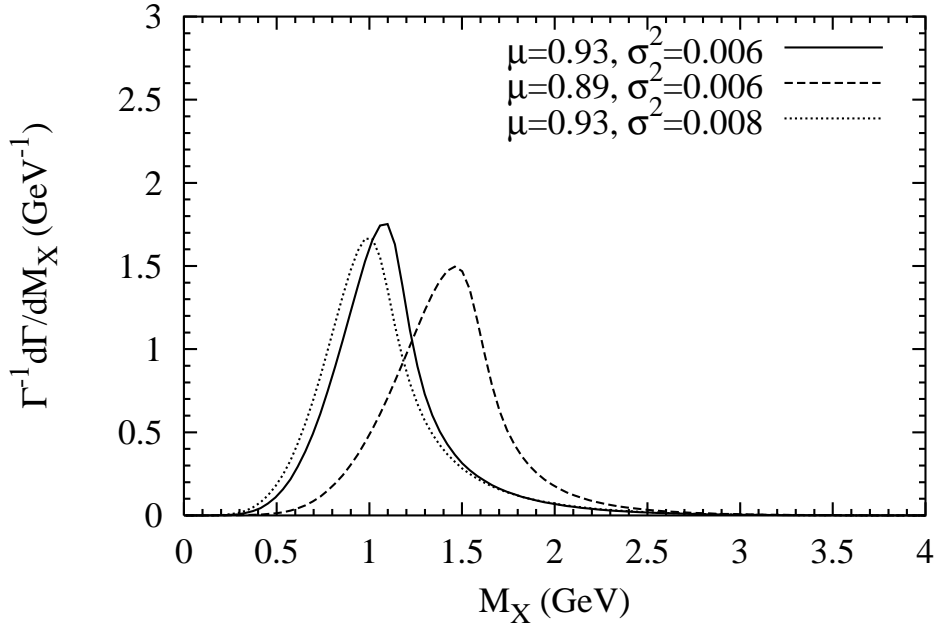}}
\tighten{
\caption[tau1]{Dependence of the charmless hadronic invariant mass spectrum 
on the mean value and the variance of the distribution function. 
The strong coupling constant is taken to be $\alpha_s=0.25$. } 
\label{fig:mx_u2} }
\end{figure}

We explore next the sensitivity of the spectrum to the form of 
the distribution function.
The mean value and the variance of it vary in the ranges specified
in Eqs.~(\ref{eq:constrain1}) and (\ref{eq:constrain2}) inferred from
the sum rules (\ref{eq:sum1}) and (\ref{eq:sum2}).
We show in Fig.~\ref{fig:mx_u2} 
the variation of the hadronic invariant mass spectrum due to the changes in 
the mean value $\mu$ and the variance $\sigma^2$ of the distribution function. 
The position of the maximum for the spectrum is a sensitive function 
of the mean value $\mu$. The sensitivity of the spectrum to the variance 
$\sigma^2$ is seen to be relatively weak.

\begin{figure}[pthb]
\centerline{\epsfysize=9truecm \epsfbox{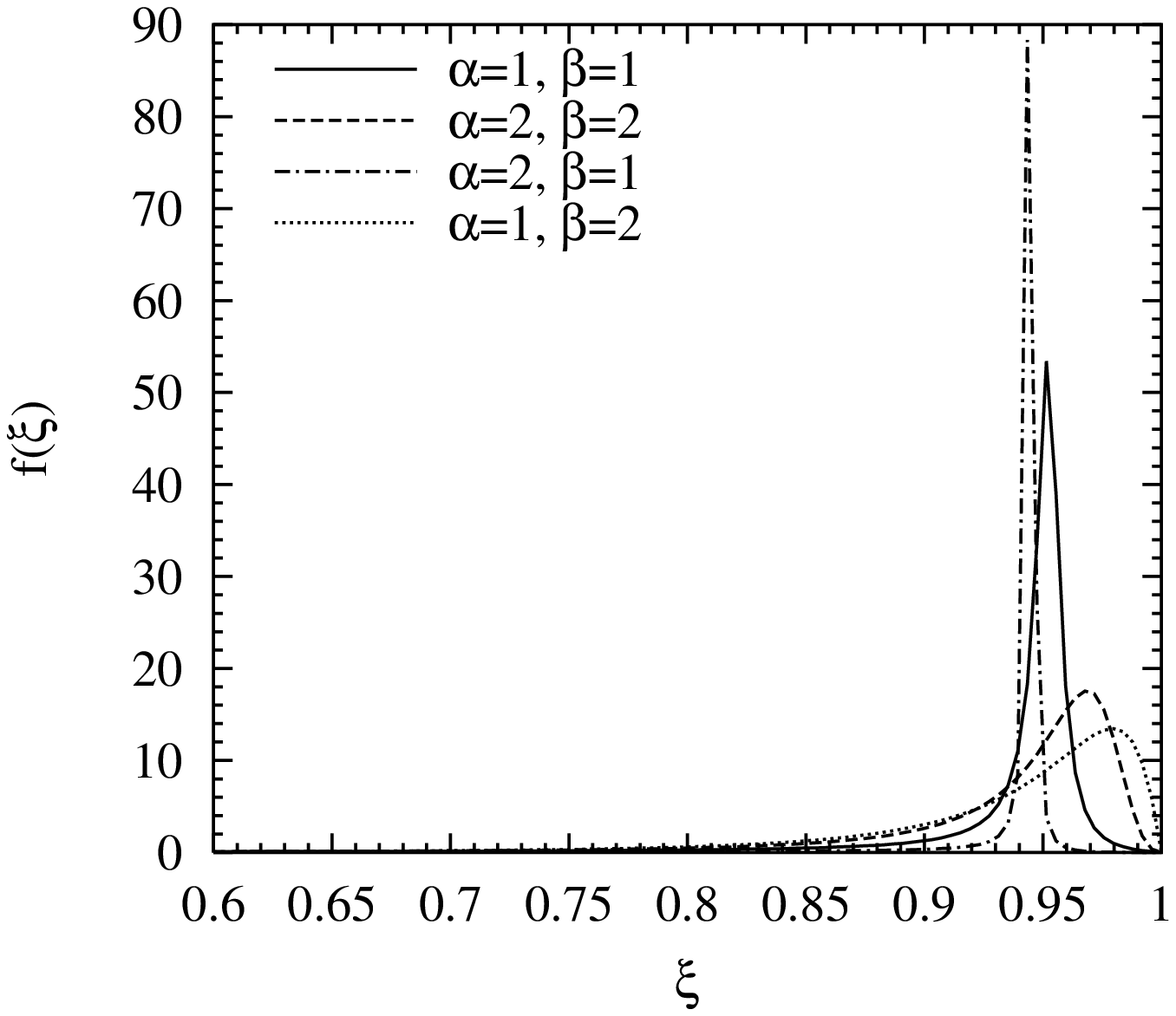}}
\tighten{
\caption[tau1]{The distribution functions from the parametrization 
(\ref{eq:ansatz})
for four different pairs of the parameters $\alpha$ and $\beta$. 
All the distribution functions give $\mu=0.93$ and $\sigma^2=0.006$. } 
\label{fig:form} }
\vspace{1truecm}
\centerline{\epsfysize=9truecm \epsfbox{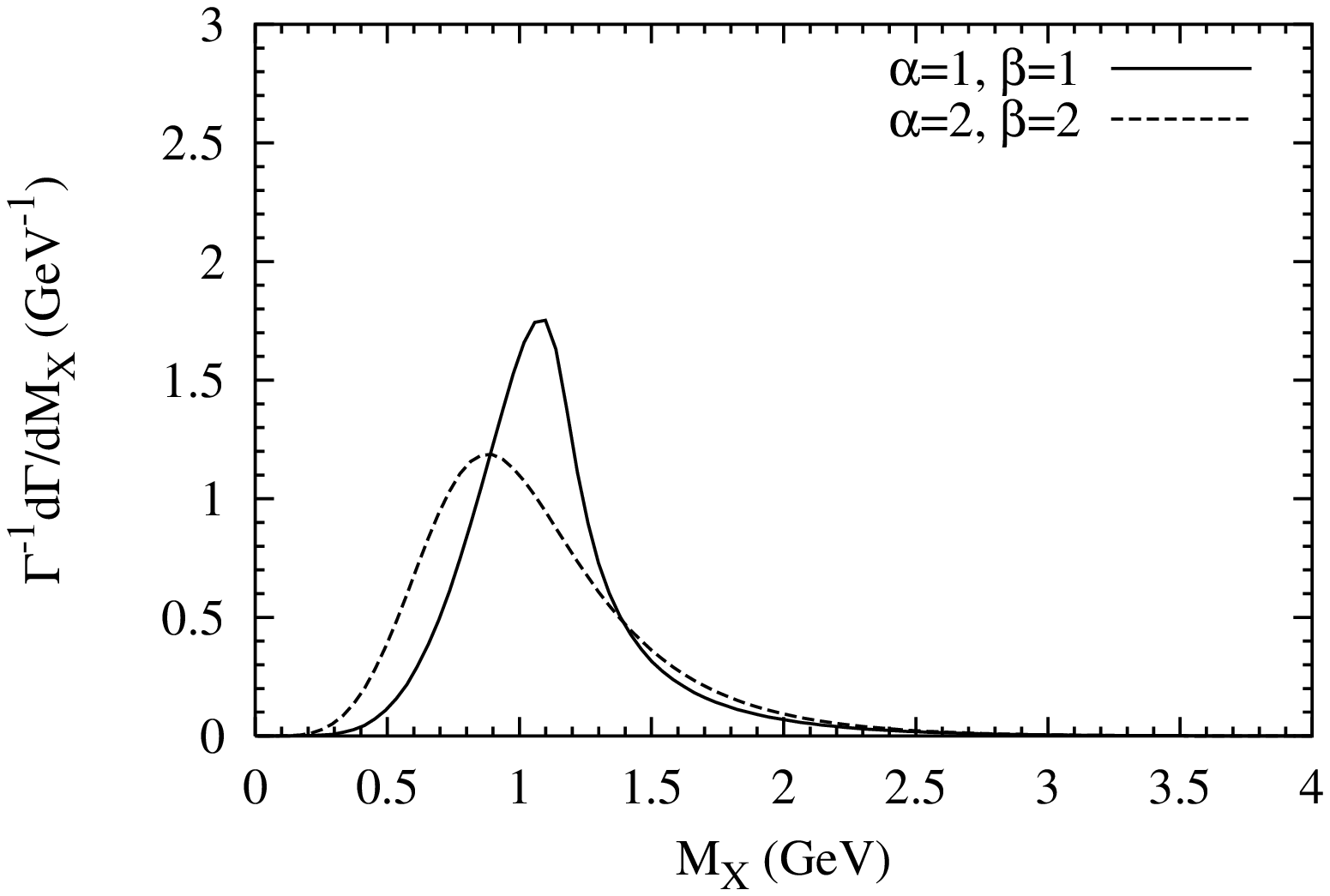}}
\tighten{
\caption[tau1]{Comparison of the
charmless hadronic invariant mass spectra in two distribution 
functions. The mean value and the variance of the distribution functions
are fixed to be $\mu=0.93$ and $\sigma^2=0.006$. We take $\alpha_s=0.25$.} 
\label{fig:mx_u3} }
\end{figure}

Furthermore, to get an idea of the sensitivity of the spectrum to the form of
the distribution function when keeping the mean value and the variance of 
it fixed, we change the shape of the distribution function by varying the 
values of the two additional parameters $\alpha$ and $\beta$ in the 
parametrization (\ref{eq:ansatz}) (for given values of $\alpha$ and $\beta$,
the parameters $a$ and $b$ are fixed by the mean value and the variance).
In Fig.~\ref{fig:form}  
we plot the distribution functions for four different
choices of the parameters $\alpha$ and $\beta$ in (\ref{eq:ansatz})
that give the same mean value $\mu=0.93$ and variance $\sigma^2=0.006$.  
The spectra calculated using the two very different distribution functions 
with $\alpha=1$ and $\beta=1$ (solid curve), 
$\alpha=2$ and $\beta=2$ (dashed curve), respectively,
at fixed mean value and variance are shown in Fig.~\ref{fig:mx_u3}. 
It is seen that the hadronic invariant mass spectrum relies strongly on
the form of the distribution function, even when the mean value and the
variance of it are kept fixed.

\section{Partial (Total) Decay Widths and Determination of 
$|V_{\lowercase{ub}}|$}

Using the charmless hadronic invariant mass spectrum, the semileptonic
branching fraction $\Delta{\cal B}_u(M_X^{\rm cut})$ for the charmless
semileptonic $B$ decay with the hadronic final states up to a given maximum
invariant mass $M_X^{\rm cut}$ can be measured.
Together with the measured lifetime of the $B$ meson, $\tau_B$, and
the theoretically calculated decay width defined in Eq.~(\ref{eq:partial})
up to the CKM factor $|V_{ub}|^2$, this determines $|V_{ub}|$:
\begin{equation}
|V_{ub}|^2=\frac{\Delta{\cal B}_u(M_X^{\rm cut})}
{\tau_B\gamma_u(M_X^{\rm cut})}\, ,
\label{eq:vub}
\end{equation}
where $\gamma_u(M_X^{\rm cut})$ is defined by $\Gamma(M_X^{\rm cut})=
|V_{ub}|^2\gamma_u(M_X^{\rm cut})$.

\begin{figure}[t]
\centerline{\epsfysize=9truecm \epsfbox{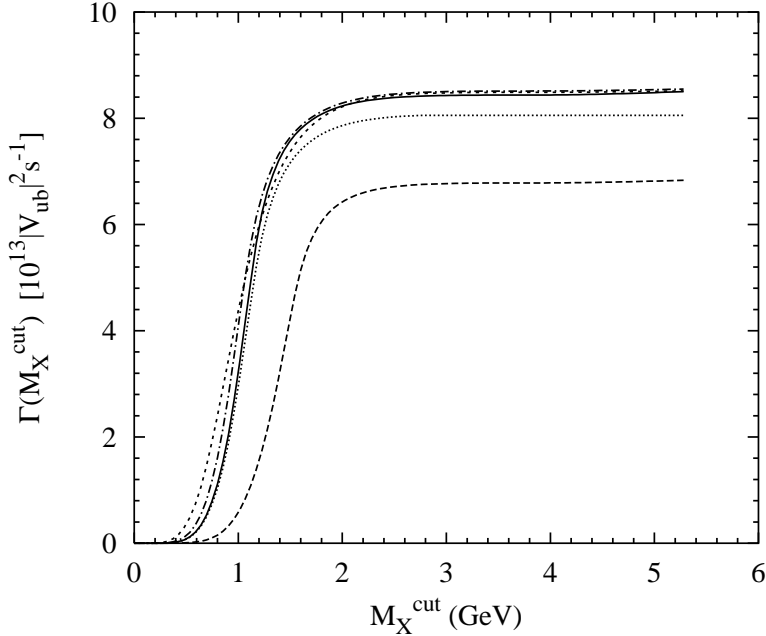}}
\tighten{
\caption[tau1]{The semileptonic decay width defined in Eq.~(\ref{eq:partial})
for $B\to X_u\ell\nu$ as a function of the hadronic invariant mass cutoff. 
The dependence on the mean value and the variance of
the distribution function is shown using $\alpha_s=0.25$ and 
the parametrization (\ref{eq:ansatz}) with $\alpha=\beta=1$ and three pairs of
the mean value and the variance: 
$\mu=0.93$ and $\sigma^2=0.006$ (solid curve),
$\mu=0.89$ and $\sigma^2=0.006$ (long-dashed curve),
$\mu=0.93$ and $\sigma^2=0.008$ (dot-dashed curve).
The dependence on the form of the distribution function at fixed mean value
and variance is shown by the solid and short-dashed curves corresponding to
the same values of $\alpha_s$ and of the mean value and the variance,
but $\alpha=\beta=2$ for the short-dashed curve.
The dependence on the strong coupling constant is illustrated by the
solid and dotted curves obtained from the same parameters except 
$\alpha_s=0.30$ for the dotted curve.}   
\label{fig:wid_cut} }
\end{figure}

We investigate the theoretical uncertainties in this determination of
$|V_{ub}|$. The semileptonic decay width $\Gamma(M_X^{\rm cut})$
is displayed in Fig.~\ref{fig:wid_cut} 
as a function of the hadronic invariant mass cutoff, $M_X^{\rm cut}$. 
For $M_X^{\rm cut}=M$, $\Gamma(M_X^{\rm cut})$ is the total semileptonic
decay width for $B\to X_u\ell\nu$. The solid and dotted curves
correspond to two different values of $\alpha_s$: $\alpha_s=0.25$ (solid),
$\alpha_s=0.30$ (dotted), and the other parameters are identical.
We find that a reasonable variation of the value of $\alpha_s$ 
leads to a small change in the value of the decay 
width $\Gamma(M_X^{\rm cut})$.

In order to explore the impact of the form of the distribution function,
as in the last section, we first study the sensitivities of
$\Gamma(M_X^{\rm cut})$ to the mean value and the variance of the distribution
function by using the parametrization (\ref{eq:ansatz}) setting 
$\alpha=\beta=1$.
We vary the mean value and the variance in the ranges of 
Eqs.~(\ref{eq:constrain1}) and (\ref{eq:constrain2}). 
The resulting decay widths are shown in Fig.~\ref{fig:wid_cut} 
corresponding to 
$\mu=0.93$ and $\sigma^2=0.006$ (solid curve),
$\mu=0.89$ and $\sigma^2=0.006$ (long-dashed curve),
$\mu=0.93$ and $\sigma^2=0.008$ (dot-dashed curve),
with $\alpha_s=0.25$ held fixed.
Then we change the form of the distribution function by changing
the values of the two additional parameters $\alpha$ and $\beta$ in
Eq.~(\ref{eq:ansatz}) in order to
examine the further sensitivity of $\Gamma(M_X^{\rm cut})$ to the distribution
function when keeping the mean value and the variance of it fixed.
The results are also shown in Fig.~\ref{fig:wid_cut} for $\alpha=\beta=1$
(solid curve), $\alpha=\beta=2$ (short-dashed curve), with $\mu=0.93,
\sigma^2=0.006$, and $\alpha_s=0.25$ held fixed for both curves.
We find that for the cut $M_X^{\rm cut}$ less than about 1.2 GeV, the 
partial decay width $\Gamma(M_X^{\rm cut})$ is rather sensitive to the form 
of the distribution function, 
whereas for $M_X^{\rm cut}$ greater than about 1.2 GeV, the decay     
width $\Gamma(M_X^{\rm cut})$ (including the total semileptonic decay width
for $B\to X_u\ell\nu$ decays) is sensitive essentially only to the mean value 
of the distribution function, which is calculable from HQET with the results
given in Eqs.~(\ref{eq:sum1}) and (\ref{eq:constrain1}). 
These imply that the theoretical uncertainties in 
the determination of $|V_{ub}|$ are under control for the value of 
the hadronic invariant mass cutoff greater than about 1.2 GeV, but very large 
for the value of the cutoff less than 1.2 GeV or so.

Adding the errors on the semileptonic decay width due to the value of 
$\alpha_s$ and the form of the distribution function linearly to be
conservative,
we estimate the theoretical errors in the determination of $|V_{ub}|$.
Without any cut, the theoretical error in determining $|V_{ub}|$ from the
total charmless semileptonic decay width is about $13\%$.  Applying
a cut on the hadronic invariant mass, the theoretical errors in determining
$|V_{ub}|$ from the hadronic invariant mass spectrum are approximately 
$16\%$ for $M_X^{\rm cut}=1.8$ GeV, $26\%$ for $M_X^{\rm cut}=1.5$ GeV,
and $44\%$ for $M_X^{\rm cut}=1.2$ GeV, respectively.

\section{Conclusion and Discussion}

We have presented an analysis of the hadronic invariant mass spectrum and
studied the decay width as a function of the invariant mass cut for 
inclusive charmless semileptonic $B$ decays. 
An approach based on the light-cone expansion
and the heavy quark effective theory has been exploited. 
Nonperturbative QCD effects
in the processes are contained in a distribution function and several 
important properties of it are known from QCD. However, one needs a 
parametrization of the distribution function incorporating the known properties
in order to make quantitative predictions, since it cannot yet been completely
determined from QCD. For our analysis, we employ the distribution function
parametrization given in Eq.~(\ref{eq:ansatz}).
We have examined the sensitivities of the hadronic invariant mass spectrum
and the decay width with a hadronic invariant mass cut to the parameters of
the theory, namely the strong coupling constant and the form of the
distribution function.  

It is shown that a reasonable variation of the value of $\alpha_s$ hardly 
affects the charmless hadronic invariant mass spectrum. The dependence
of the decay width with a hadronic invariant mass cut on the value of
$\alpha_s$ is small. The main theoretical
uncertainties arise from the form of the distribution function.

We found that the charmless hadronic invariant mass spectrum exhibits a
significant sensitivity to the form of the distribution function. Thus,
if experimentally feasible, the measured hadronic invariant mass spectrum 
will impose strong constraints on the shape of the distribution function.

We have also studied the sensitivity of the semileptonic decay width to a given
maximum hadronic invariant mass to the form of the distribution function.
For the hadronic invariant mass cutoff greater than about 1.2 GeV, the
value of the decay width is sensitive essentially only to the mean value
of the distribution function whose value is known from HQET, and thereby
the result is nearly
model independent and the uncertainties in the calculation
of the decay width are under control. However, the theoretical
uncertainties in the partial decay width with a hadronic invariant mass
cutoff less than 1.2 GeV are very large, obscuring its usefulness for
determining $|V_{ub}|$, since it is rather sensitive to the form of the
distribution function.

We have observed the feature that for $M_X^{\rm cut}\geq M_D$ the decay
width tends to show only little dependence on the invariant mass cut
(see Fig.~5) and most of the $b\to u$ events lie below $M_X=M_D$ (see also
Figs.~1, 2, 4).

We have estimated the theoretical errors in determining $|V_{ub}|$ in our
approach. At the present time, the theoretical error on $|V_{ub}|$ from 
the total semileptonic decay width for the $B\to X_u\ell\nu$ decays
is about $13\%$. Applying a hadronic invariant mass cutoff, the theoretical
errors on $|V_{ub}|$ are approximately $16\%$ for $M_X^{\rm cut}=1.8$ GeV,
$26\%$ for $M_X^{\rm cut}=1.5$ GeV, and $44\%$ for $M_X^{\rm cut}=1.2$ GeV,
respectively. 
The higher the hadronic invariant mass cutoff can be experimentally made to 
be, the smaller
the theoretical error on $|V_{ub}|$; the smallest theoretical error on 
$|V_{ub}|$ would be achieved if the total charmless semileptoinc decay 
width can be measured\footnote{An attempt has been made \cite{aleph}
by the ALEPH Collaboration to measure the inclusive charmless semileptonic 
branching fraction of $B$ hadrons.}.

Besides the hadronic invariant mass cutoff, the mean value of the distribution
function is another key quantity for precise determination of $|V_{ub}|$,
which requires as precise knowledge of the mean value as possible.
Based on this analysis and the previous one \cite{jp1}, 
a detailed fit to the measured
charged-lepton energy spectrum and/or the hadronic invariant mass spectrum
can be used to extract the parameters (especially, the mean value of the 
distribution function) experimentally. This procedure allows to reduce the 
uncertainties in the determination of $|V_{ub}|$. 
Future measurements of the distribution function and
continuing efforts in calculations of hadronic matrix elements 
that govern $B$ decays should enable to further reduce the uncertainties.

\acknowledgments
I would like to thank Emmanuel Paschos for discussions.
I also wish to thank Thomas Mannel and the Institut f\"ur Theoretische 
Teilchenphysik at the Universit\"at Karlsruhe for the warm hospitality.
This work has been supported in part by BMBF.

{\tighten

} 


\begin{references}

\bibitem{cleo1} CLEO Collaboration, R. Fulton {\it et al.}, 
Phys. Rev. Lett. {\bf 64}, 16 (1990);
J. Bartelt {\it et al.}, {\it ibid.} {\bf 71}, 4111 (1993).

\bibitem{argus} ARGUS Collaboration, H. Albrecht {\it et al.}, 
Phys. Lett. B {\bf 234}, 409 (1990); {\bf 255}, 297 (1991).

\bibitem{aleph} ALEPH Collaboration, D. Buskulic {\it et al.},
contributed paper PA05-59 to the 28th International Conference on
High Energy Physics, Warsaw, Poland, 1996; H. Kroha,
in {\it Proceedings of the 28th International Conference on High
Energy Physics}, Warsaw, Poland, 1996, edited by Z. Ajduk and
A.K. Wroblewski (World Scientific, Singapore, 1997), p.~1182.

\bibitem{richman} For recent reviews, see 
J.D. Richman and P.R. Burchat,
Rev. Mod. Phys. {\bf 67}, 893 (1995);
T.E. Browder and K. Honscheid, Progress in Nuclear and Particle
Physics {\bf 35}, 81 (1995);
J.R. Patterson, in {\it Proceedings of the 
28th International Conference
on High Energy Physics}, Warsaw, Poland, 1996, edited by Z. Ajduk and
A.K. Wroblewski (World Scientific, Singapore, 1997), p.~871.

\bibitem{jp1} C.H. Jin and E.A. Paschos, DO-TH 96/22, LMU 03/97,
hep-ph/9704405, to appear in Z. Phys. C.

\bibitem{cleo2} CLEO Collaboration, J.P. Alexander {\it et al.},
Phys. Rev. Lett. {\bf 77}, 5000 (1996).

\bibitem{gib} L.K. Gibbons, in {\it Proceedings of the 
28th International Conference
on High Energy Physics}, Warsaw, Poland, 1996, edited by Z. Ajduk and
A.K. Wroblewski (World Scientific, Singapore, 1997), p.~183.  

\bibitem{rev} For recent reviews, see  
P. Ball, to appear in {\it Proceedings of the 7th International Symposium 
on Heavy Flavor Physics},
Santa Barbara, USA, 1997, hep-ph/9709407;
C.T. Sachrajda, to appear in {\it Proceedings of the XVIII International 
Symposium on Lepton-Photon Interactions}, Hamburg, Germany, 1997, 
hep-ph/9711386.  

\bibitem{bar} A. Bareiss and E.A. Paschos, 
Nucl. Phys. {\bf B327}, 353 (1989); 
C.H. Jin, W.F. Palmer, and E.A. Paschos, DO-TH 93/21, OHSTPY-HEP-T-93-011
(1993) (unpublished);
Phys. Lett. B {\bf 329}, 364 (1994);
V. Barger, C.S. Kim, and R.J.N. Phillips, {\it ibid.}
{\bf 235}, 187 (1990); {\bf 251}, 629 (1990); C.S. Kim, D.S. Hwang, P. Ko,
and W. Namgung, Nucl. Phys. B (Proc. Suppl.) {\bf 37A}, 69 (1994);
Phys. Rev. D {\bf 50}, 5762 (1994).

\bibitem{sev} A.F. Falk, Z. Ligeti, and M.B. Wise, 
Phys. Lett. B {\bf 406}, 225 (1997);
I. Bigi, R.D. Dikeman, and N. Uraltsev, hep-ph/9706520;
R.D. Dikeman and N. Uraltsev, hep-ph/9703437;
C.S. Kim, Phys. Lett. B {\bf 414}, 347 (1997).

\bibitem{jp} C.H. Jin and E.A. Paschos, in {\it Proceedings of the 
International 
Symposium on Heavy Flavor and Electroweak Theory}, Beijing, China, 
1995, edited by C.H. Chang and C.S. Huang (World Scientific, Singapore,
1996), p.132; DO-TH 95/07, hep-ph/9504375.
\bibitem{jin} C.H. Jin, Phys. Rev. D {\bf 56}, 2928 (1997).

\bibitem{jin1} C.H. Jin, Phys. Rev. D {\bf 56}, 7267 (1997).  

\bibitem{drell} P.S. Drell, to appear in {\it Proceedings of the XVIII 
International Symposium
on Lepton-Photon Interactions}, Hamburg, Germany, 1997, hep-ex/9711020.  

\bibitem{hqe} J. Chay, H. Georgi, and B. Grinstein, Phys. Lett. B {\bf 247},
399 (1990);
I.I. Bigi, N.G. Uraltsev, and A.I. Vainshtein, {\it ibid.} {\bf 293}, 
430 (1992); {\bf 297}, 477(E) (1993);
I.I. Bigi, M.A. Shifman,
N.G. Uraltsev, and A.I. Vainshtein, Phys. Rev. Lett. {\bf 71}, 496 (1993);
A.V. Manohar and M.B. Wise, Phys. Rev. D {\bf 49}, 1310 (1994);
B. Blok, L. Koyrakh, M.A. Shifman, and A.I. Vainshtein, 
{\it ibid.} {\bf 49}, 3356 (1994); {\bf 50}, 3572(E) (1994).

\bibitem{PDG} Particle Data Group, R.M. Barnett {\it et al.}, Phys. Rev.
D {\bf 54}, 1 (1996).

\bibitem{ball} P. Ball and V. Braun, Phys. Rev. D {\bf 49}, 2472 (1994).

\bibitem{greub} C. Greub and S.-J. Rey, Phys. Rev. D {\bf 56}, 4250 (1997),
and references therein.  

\end{references}
\end{document}